\documentclass[12pt,preprint]{emulateapj}
\usepackage{url,color}

\begin{document}

\title{Testing the Binary Trigger Hypothesis in FUors}

\author{
Joel D. Green\altaffilmark{1,2},
Adam L. Kraus\altaffilmark{2},
Aaron C. Rizzuto\altaffilmark{2},
Michael J. Ireland\altaffilmark{3},
Trent J. Dupuy\altaffilmark{2},
Andrew W. Mann\altaffilmark{2},
\& Rajika Kuruwita\altaffilmark{3}
}

\altaffiltext{1}{Space Telescope Science Institute, Baltimore, MD, USA}
\altaffiltext{2}{Dept of Astronomy, The University of Texas at Austin, Austin, TX 78712, USA}
\altaffiltext{3}{Research School for Astronomy \& Astrophysics, Australia National University, Canberra ACT 2611, Australia}

\begin{abstract}

We present observations of three FU Orionis objects (hereafter, FUors) with nonredundant aperture-mask interferometry (NRM) at 1.59 $\mu$m and 2.12 $\mu$m that probe for binary companions on the scale of the protoplanetary disk that feeds their accretion outbursts. We do not identify any companions to V1515 Cyg or HBC 722, but we do resolve a close binary companion to V1057 Cyg that is at the diffraction limit ($\rho = 58.3 \pm 1.4$ mas or $30 \pm 5$ AU) and currently much fainter than the outbursting star ($\Delta K' = 3.34 \pm 0.10$ mag). Given the flux excess of the outbursting star, we estimate that the mass of the companion ($M \sim 0.25 M_{\odot}$) is similar to or slightly below that of the FUor itself, and therefore it resembles a typical T Tauri binary system. Our observations only achieve contrast limits of $\Delta K' \sim 4$ mag, and hence we are only sensitive to companions that were near or above the pre-outburst luminosity of the FUors.  It remains plausible that FUor outbursts could be tied to the presence of a close binary companion. However, we argue from the system geometry and mass reservoir considerations that these outbursts are not directly tied to the orbital period (i.e., occurring at periastron passage), but instead must only occur infrequently.

\end{abstract}

\keywords{}

\section{Introduction}

The low mass pre-main sequence star FU Orionis (hereafter, FU Ori) brightened by $\Delta B = 6$ mag in 1936 \citep{herbig77}, and subsequently has declined slowly ($\sim$ 0.013 mag/yr; \citealt{kenyon00}) to the present day. This outburst established a prototype class of variable star (FU Orionis objects, or ``FUors''; \citealt{hartmann96a,hartmann98}) that $\sim$10 other stars have since matched. \citet{paczynski76} first proposed that FUors are the result of a sudden cataclysmic accretion of material from a reservoir that had built up in the circumstellar disk surrounding a young stellar object. In this model, the accretion rate rises from the typical rate for a T Tauri star ($\dot{M} \lesssim 10^{-7} M_{\odot}$ yr$^{-1}$) up to $\dot{M} \sim 10^{-4} M_{\odot}$ yr$^{-1}$, and then decays 
over an e-fold time of $\sim$ 10-100 yr \citep{lin85,hartmann85,bell94}.   Over the entire outburst the star could accrete $\sim$ 0.01 M$_{\odot}$ of material, roughly the entire mass of the Minimum Mass Solar Nebula or a typical T Tauri disk \citep{andrews05}.  On average, FUors should occur (or recur) 5-10 times per star formed in the local region of the Milky Way \citep{hartmann96a}.

FUors offer a potential counterpoint to the recent discovery of the ``luminosity problem''. In the family of isothermal collapse/accretion models \citep{shu77}, most young stars accrete at modest rates that gradually diminish as the system evolves, the circumstellar envelope thins, and the accretion disk accretes onto the central star, agglomerates in disk regions, or forms protoplanetary objects. However, the observed steady-state accretion luminosities of low mass young stars and protostars appear to be insufficient to explain the total mass accretion required to form the central objects (e.g., \citealt{kenyon90,dunham12a}). One part of the solution may be rapid early accretion prior to formation of a disk (e.g. \citealt{federrath10}), but once a massive disk is formed, FUors provide a potential solution. The rare dramatic outbursts of FUors could account for the missing mass accretion if they represent a stage that all protostars occupy for a small fraction of their lifetime. These bursts also modify the protoplanetary disk chemistry and require a very different model than simple magnetospheric accretion \citep[e.g.][]{green06, quanz07a, zhu07}.  If FUors represent a short-duration stage that all young stars undergo, then understanding their properties is vital to models of planet formation and the evolution of protoplanetary gas-rich disks.  

Understanding the triggering mechanism to FUors is key to differentiating between models of ``inside-out'' vs. ``outside-in'' collapse.  The former begins with an event very near the inner disk edge (whether the accretion of a planetary mass worth of material or an overflow of the magnetospheric radius by material spiraling in), and the latter begins with an event farther out, possibly up to several AU, where the lower temperatures, and therefore lower ionization fraction, render the magneto rotational instability (MRI) inefficient as an accretion mechanism \citep[e.g.][]{martin12}.  A third option that has been posited is an external perturber -- the interaction of the disk with an exterior massive object that alters the balance of accretion and triggers the cascading flow.  Indeed, the archetypal FUor, FU Ori, is itself a binary system \citep{wang04} and one of the most dramatic bursts observed \citep{herbig77,audard14}.  If these bursts only occurred in binary systems, we would expect a significant chemical segregation between binary and single systems, as only the former would have experienced this temperature increase in the inner disk during the epoch of planet formation.  If instead most or all systems undergo these bursts, they may be sufficient to resolve the luminosity problem and need to be factored into models including the initial conditions for planet formation.

In this letter, we report nonredundant aperture-mask interferometry observations of three FUors (V1057 Cyg, V1515 Cyg, and HBC 722) in the Cygnus star-forming complex, and the discovery that V1057 Cyg has a close ($\rho \sim 30$ AU) binary companion that is likely similar in mass to the outbursting star.

\section{Sample}

\begin{deluxetable*}{lrrrllll}
 \tabletypesize{\tiny}
 \tablewidth{0pt}
 \tablecaption{Sample}
 \tablehead{
 \colhead{2MASS J} & \colhead{Other} & \colhead{$V$ (mag)} & \colhead{$K$ (mag)} & \colhead{Dist. (pc)} &\colhead{Burst Year} & \colhead{SED Class} & \colhead{Refs}}
 \\
 \colhead{} & \colhead{Name} & \colhead{(mag)} & \colhead{(mag)} & \colhead{(pc)} & \colhead{} & \colhead{}
   
 \startdata

20585371+4415283 & V1057 Cyg & $12.43 \pm 0.03$ & $6.227 \pm 0.017$ & $600 \pm 100$ & 1970 & FS & Hartmann \& Kenyon 1996  \\
20234802+4212257 & V1515 Cyg & $13.44 \pm 0.04$ & $7.378 \pm 0.021$ & $1000 \pm 200$ & 1950 & FS/II & Millan-Gabet et al. 2006 \\
20581702+4353433 & HBC 722 & $13.38 \pm 0.05$ & $7.90 \pm 0.02$ & $600 \pm 100$ & 2010 &  II & H. Sung et al. (2013)\\

\enddata
\tablecomments{V magnitudes are computed from the mean and standard deviation of all values reported to the AAVSO between 2015 July 10 and 2015 August 1. The K magnitudes are computed from 2MASS $K_s$ (V1057 Cyg and V1515 Cyg) or from \citet{sung13} (HBC 722), assuming the K-band flux has leveled out since 2013, as it has in the optical bands (source: AAVSO).\\}
 \end{deluxetable*}
 
Our sample (summarized in Table 1) includes two classical FUors that outburst in 1950 (V1515 Cyg) and in 1970 (V1057 Cyg; \citealt{herbig77}), and one of the most recent confirmed FUors that outburst in 2010 (HBC 722/V2493 Cyg; \citealt{semkov10a,semkov10b}).  V1057 Cyg and HBC 722 are the only FUors with pre-outburst spectra; both indicate young (T Tauri-like) stars \citep{herbig77,miller11,kospal11}.  All three show significant reddening ($A_V$ $\sim$ 3; \citealt{audard14}).  V1057 Cyg and V1515 Cyg exhibit flat IR spectral energy distributions on the (observationally defined) Class I/II boundary, with weak but pristine silicate emission features at 10 and 20 $\mu$m \citep{green06, quanz07a}.  HBC 722 burst after the end of the cryogenic lifetime of Spitzer-IRS but has been observed with SOFIA/FORCAST and Herschel, and appears to have an SED more typical of a Class II source \citep{green11b,green13c}.  HBC 722 and V1057 Cyg are both located in the North America Nebula; V1515 Cyg is more distant in the Cygnus complex.  Modeling of the SEDs indicate central star masses of 0.3-0.5 M$_{\odot}$. 

We computed contemporaneous $V$ magnitudes for each target from the mean and standard deviation of all values reported to the AAVSO (Kafka 2016) by amateur astronomers between 2015 July 10 and 2015 August 1. In order to estimate contemporaneous $K$ magnitudes, which are not monitored by the AAVSO, we used 2MASS $K_s$ values and assumed that the variation from 2000 until present was linearly proportional to the variation in optical bands, for V1057 Cyg and V1515 Cyg.  For HBC 722, \citet[e.g.][Fig. 2]{sung13} monitored this source regularly throughout the outburst.    None of the sources showed significant mid-IR (WISE bands 1 and 2) variability in 2010-2011 \citep{cutri12,cutri13}. Extrapolating their data from June 2013, the optical brightness of the source has not varied significantly, and thus we take the $K_s$ mag as unchanged from 2013 \citep[][Fig. 1]{sung13}.  The HBC 722 light curve illustrates the similarity of variability in all of the optical/near-IR bands; $K_s$ varies approximately as two-thirds of the variation (in magnitudes) in V or B.  Using this relation, the $\Delta B = 6$ mag outburst of V1057 Cyg was likely $\Delta K \sim 4$ mag in total amplitude, and has now declined to an excess of $\Delta K \sim 2.7$ mag.

\section{Observations and Data Analysis}

\begin{figure}
\epsscale{1.0}
\includegraphics[scale=0.5,trim={0 0 0 0},clip]{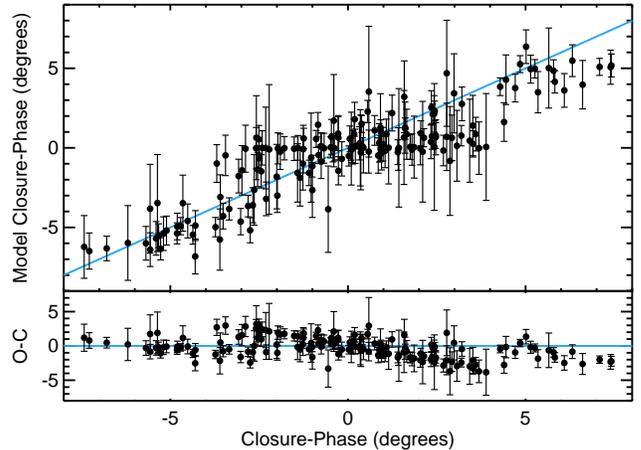}
\figcaption{\label{fig:model} {\bf Top:} For our $K'$ observations of V1057 Cyg AB, the observed closure phases plotted against the corresponding values of the best-fit model. The companion is near the detection limits, so there significant scatter about the 1:1 line, but it is detected at $>$99.9\% confidence.  {\bf Bottom:} The fit residuals. }\end{figure}

\begin{figure*}
\epsscale{1.0}
\includegraphics[scale=0.95,trim={0 0 0 0},clip]{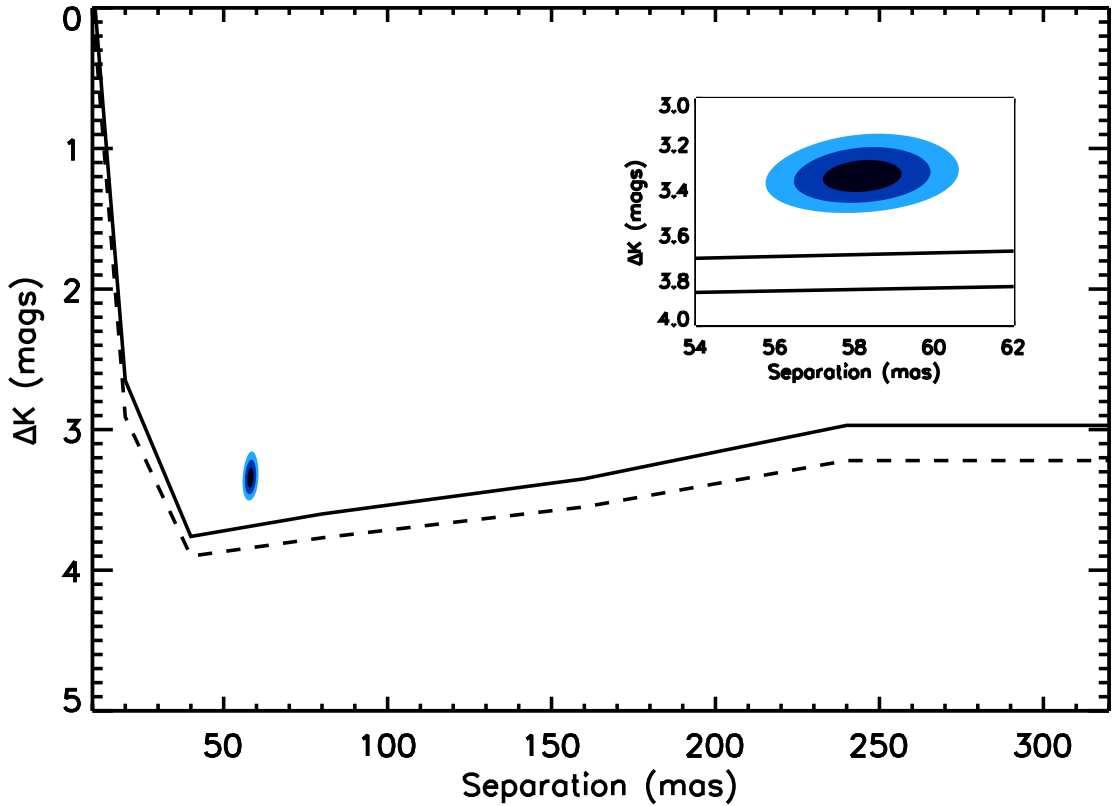}
\figcaption{\label{fig:fit} Companion confidence interval and detection limits for V1057 Cyg AB, in terms of the contrast $\Delta K'$ and projected angular separation $\rho$. The ellipses in the main figure and inset show the 1, 3, and 6 sigma joint confidence intervals on the contrast and projected separation for V1057 Cyg B with respect to the outbursting star, and the solid and dashed lines show the 99.9\% and 99\% detection limits for additional companions.}
\end{figure*}

 \begin{deluxetable}{lccccc}
\tabletypesize{\footnotesize}
\tablewidth{0pt}
\tablecaption{Keck/NIRC2 NRM Detection Limits}
\tablehead{
\colhead{Name} & \colhead{Filter} &  \colhead{$N_{frames}$} & \multicolumn{3}{c}{Limits (mag) at} 
\\
\colhead{} & \colhead{} & \colhead{} &  \colhead{10 mas} & \colhead{20 mas} & \colhead{40 mas}
}
\startdata
HBC 722 & CH4S & 16 & 0.21 & 3.15 & 3.27 \\
HBC 722 & K'   & 16 & 0.00 & 2.93 & 4.03 \\
V1057 Cyg & CH4S & 16 & 0.45 & 3.39 & 3.47 \\
V1057 Cyg & K'   & 17 & 0.00 & 2.59 & 3.63 \\
V1515 Cyg & CH4S & 8 & 0.00 & 2.47 & 2.53 \\
V1515 Cyg & K'   & 9 & 0.00 & 2.50 & 3.69 \\
\enddata
\end{deluxetable}

Non-redundant aperture mask interferometry (NRM) is now a well-established technique to achieve the full diffraction limit of a single telescope \citep{Nakajima:1989zl,Tuthill:2000ge,Tuthill:2006kq,Ireland:2013lr}, superior even to standard AO imaging. The core innovation of NRM is to resample the telescope's single aperture into a sparse interferometric array.  NRM allows for data analysis using interferometric techniques (such as closure-phase analysis) that calibrate out the phase errors that limit traditional astronomical imaging by inducing speckle noise. As we described in \citet{Kraus:2008zr,Kraus:2011tg}, NRM observations can yield contrasts as deep as $\Delta K \sim 4$ at 1/2 $\lambda /D$ with observations of $\sim$15 minutes, and we have used the technique to identify dozens of binary companions that fall inside the detection limits of traditional imaging surveys, although FUors are among the youngest, early stage protostellar objects to which NRM has been applied. More detailed discussions of the benefits and limitations of NRM, as well as typical observing strategies, can be found in \citet{Kraus:2008zr} and in \citet{Readhead:1988so,Tuthill:2000ge,Martinache:2007ru,Ireland:2008yq}.

We observed the targets with the Keck-II telescope using laser guide star adaptive optics and NRM on 2015 July 22 ($MJD=57225.6$). All observations were taken in vertical angle mode using the facility adaptive optics imager NIRC2, which has a 9-hole aperture mask installed in a cold filter wheel near the pupil stop. Our standard observing mode uses 20s integrations in a 512$\times$512 pixel subarray, each taken with a single 20s coadd and 64 Fowler samples. A typical visit to a target consists of 8 frames each in the $K'$ ($\lambda = 2.124 \mu$m) and $CH4S$ ($\lambda = 1.592 \mu$m) filters, though in some cases a ninth frame was taken if one visually appeared to have lower image quality. We obtained two such visits for HBC 722 and V1057 Cyg, and one visit in V1515 Cyg. The targets were observed in sequence so that they could be used as interferometric calibrators for each other.

The data analysis follows almost the same prescription as in \citet{Kraus:2008zr,Kraus:2011tg}, so we discuss here only a general background to the technique and differences from \citet{Kraus:2008zr}. The data analysis takes three broad steps: basic image analysis (flat-fielding, bad pixel removal, dark subtraction), extraction and calibration of squared visibility and closure phase, and binary model fitting. Unless fitting to close, near-equal binaries, we fit only to closure phase, as this is the quantity most robust to changes in the AO point-spread function (PSF).  We adopted the platescale and rotation of \citet{yelda10} in computing relative astrometry.

The detection limits are found using a Monte-Carlo method that simulates 10,000 random closure-phase datasets of a point source, with closure-phase errors and covariances that match those of the calibrated target data set. This routine then searches for the best fit for a companion in each randomized dataset. Over each annulus of projected separation from the primary star, the 99.9\% (3.3$\sigma$) confidence limit is set to the contrast ratio where 99.9\% of the Monte-Carlo trials have no best binary fit with a companion brighter than this limit anywhere within the annulus. In a case where a companion is detected, the errors around this two-source fit are instead used in the Monte Carlo.   The fitting procedure tests first the goodness of fit for a single point source model in the closure phases; if the fit is rejected, a fit with two point sources is then tested, assuming a point-asymmetric structure around the primary star.

\section{Results}

We have discovered one new binary companion among our three targets, a faint object in close proximity to V1057 Cyg (hereafter V1057 Cyg B) that would have been undetectable with standard adaptive optics imaging. V1057 Cyg B was detected at $>$99.9\% confidence in $K'$ and at $>$99\% confidence in $CH4S$, with consistent astrometry and photometry in independent analyses of each filter, when considering each visit separately, and when using different subsets of visits to other targets for calibration. Using all visits and all calibrating datasets, the best fit in $K'$ was for $\rho = 58.3 \pm 1.4$ mas, $\theta = 109.7 \pm 1.4 \degr$, and $\Delta m = 3.34 \pm 0.10$ mag, while the best fit in $CH4S$ was for $\rho = 53.7 \pm 2.4$ mas, $\theta = 111.6 \pm 2.3 \degr$, and $\Delta m = 3.57 \pm 0.20$ mag. The separation $\rho$, position angle $\theta$, and contrast $\Delta m$ are therefore consistent between both filters. We show a plot of the model closure phases versus the fit closure phases in each filter in Figure \ref{fig:model}.

Given the distance to the North America Nebula ($D = 600 \pm 100$ pc; \citealt{laugalys02}), the angular projected separation of V1057 Cyg B corresponds to a physical projected separation of $\rho = 30 \pm 5$ AU. The orbital period is prohibitively long to determine the orbital semimajor axis (164 years at 30 AU). However, the semimajor axis can be no less than half of the current projected separation, and is unlikely to be significantly larger given standard priors (e.g., Dupuy \& Liu 2011). Interpretation of the stellar properties is more complicated since the brightness of the primary star is significantly enhanced by accretion luminosity. The V1057 Cyg system brightened by 6 mag in the V filter in 1971 (Welin 1971; Herbig \& Harlan 1971), and remains $\sim$4 mag brighter at present. Given standard SED shapes for FUors, the outbursting star is likely $\sim$2.7 mags brighter than quiescient levels in the near-infrared (Section 2). We therefore assume that V1057 Cyg B was $\Delta K \sim$0.6 mag fainter than the outbursting star before the outburst. 

FUor hosts are generally assumed to have masses of $M \sim $0.3--0.7 $M_{\odot}$ (e.g., Hartman \& Kenyon 1996; Zhu et al. 2007), though this estimate is very uncertain when pre-outburst spectroscopy is not available. Across this mass range, the 1 Myr models of Baraffe et al. (2015) predict that a flux ratio of $\Delta K' \sim 0.6$ mag would correspond to a mass ratio of $q = $0.59--0.62, a typical value for binary companions (e.g., Raghavan et al. 2010). However, this estimate is extremely uncertain without an accurate pre-outburst primary star mass, $K$ magnitude, or measure of the (pre-outburst) disk excesses on both stars. Alternately, the observed 2MASS magnitude ($K_s = 6.23 \pm 0.02$) and our flux ratio ($\Delta K' = 3.34 \pm 0.10$) can be combined with the distance ($D = 600 \pm 100$ pc) and extinction ($A_V = 3.5 \pm 0.5$ or $A_K = 0.39$; Green et al. 2006) to yield an absolute magnitude of $M_K = 0.3^{+0.4}_{-0.3}$ for non-erupting companion. This brightness is above the top of the model grid at 1 Myr ($M > 1.4 M_{\odot}$), and the disk of the secondary would need to be 1.4 mag at $K$ in order to reduce the inferred mass to $M = 1 M_{\odot}$. We therefore suggest that while the system might be more massive than is typically assumed, no firm conclusions can be drawn yet.

We summarize the detection limits for all three targets in Table 2, and in Figure \ref{fig:fit} we show the $K'$ contrast curves for each object and the joint confidence intervals on the separation and contrast for V1057 Cyg B. Our observations achieved typical limits at $\rho = 40$ mas ($\la \lambda / D$) of $\Delta m = $3.6--4.0 mag in $K'$ and $\Delta m = $2.5--3.5 in $CH4S$, while the limits at $\rho = 20$ mas ($\la 1/2 \lambda / D$ are still $\ga$2.5 mag in all cases. Given distances of $D = 600$ pc or $D = 1000$ pc, the corresponding physical separations probed are 28/14 AU or 40/20 AU respectively, and hence fall well inside the typical radius of circumstellar disks ($R \ga 100$ AU; Andrews et al. 2005, 2007). The primary stars retain significant flux excesses of $\Delta K \ga 3$ mag from their outburst (Section 2), so these limits only encompass companions that would have had similar or brighter luminosity pre-outburst, and hence have similar or higher mass to the outbursting star. It is therefore notable that even one out of the three targets we observed does indeed have a detectable companion.

\section{Implications for the Nature of FU Orionis Objects}

A connection between binarity and outburst behavior has been examined in several studies \citep[e.g.][]{reipurth02,millan06}, further motivated by the discovery that FU Ori is itself a 0$\farcs$5 binary \citep{wang04}, since disk-binary interactions could lead to torques and large-scale warps that promote episodic accretion \citep{reipurth04}. However, the census of binarity among FUors has been hampered by their large distances ($d \ga 400$ pc) and by the presence of flux excesses as high as $\Delta K \sim 4$ due to the outburst itself; any companions will therefore be faint and near the diffraction limit. The development of NRM and interferometric techniques for the first time allow meaningful upper limits or detections of binarity in FUor systems.   If FUors occur exclusively in binary systems, the chemistry of circumstellar disks in these systems could be substantially different from single systems like the Solar System.  For example, during an FUor, the temperature at 1 AU may approach 1000 K or even higher; this becomes comparable to the temperature to form crystalline silicates or remove volatiles, which can affect habitable zone planet formation \citep{hubbard14}.  Further out in the envelope, episodic increases in temperature can cause irreversible chemical processes, such as the observed conversion of mixed CO-CO$_2$ ice into pure CO$_2$ \citep{kimhj12b}.

V1057 Cyg B, currently $\sim$ 3.34 mag fainter than the outbursting object in K', was likely of similar or slightly fainter luminosity pre-outburst and therefore represents a typical equal (low) mass young stellar object binary pair, typical of many T Tauri systems (e.g., Kraus \& Hillenbrand 2012).  If the semimajor axis is approximately equal to the current projected separation ($\rho = 30 \pm 5$ AU), then any circumstellar disk around the outbursting primary should be truncated at $a \sim 10$ AU (Artymowicz \& Lubow 1994). We therefore can extrapolate an interesting upper limit to the amount of mass available for accretion through the disk during the burst. Assuming a central star mass of $M \sim 0.3 M_{\odot}$ \citep{zhu09}, the integrated accretion luminosity of V1057 Cyg (using model parameters from \citealt{green06}) corresponds to a total accreted mass of $M \sim 0.0045 M_{\odot}$. The mass of a T Tauri disk within $r < 10$ AU, assuming a Minimum Mass Solar Nebula (MMSN; $\Sigma_{disk} \propto r^{-1.5}$; \citealt{hayashi81}) is $\sim$ 0.005 M$_{\odot}$. We therefore conclude either that the disk mass is much larger than the MMSN, or that the circumstellar disk around the outbursting star must have been nearly depleted. Given a Keplerian orbital timescale of $P \sim 300$ yr, it seems implausible that such an accretion burst could occur every orbit (i.e., from torques near periastron passage), as the inner disk could not be replenished from an envelope or circumbinary disk quickly enough. FU Ori itself has also accreted a large fraction of its disk mass in the current outburst.  The ``disk-draining'' FUors are the easiest to observe: they are more evolved (with the weakest pre-outburst infrared excess for comparison) and less embedded than sources in the earlier stages of star formation.

Furthermore, FU Ori was the target of a recent ALMA study that separated the two components \citep{hales15} and demonstrated that the remnant envelope is tenuous at best, indicating that there is no external reservoir for resupplying the disk material in repeated bursts. \citet{hales15} therefore hypothesized that the more extinguished southern component could act as the mass reservoir for resupply. The observations constrained the disk sizes to be potentially larger than for V1057 Cyg ($R_{disk} < 45$ AU), though the binary likely truncates them at $\rho \sim 30$ AU, but the mass reservoir is at least larger for FU Ori. In the case of V1057 Cyg, with a tighter binary configuration, both disks are likely to be much less massive and hence the resupply problem is even more severe.  Flattened envelope models can explain the long wavelength SED and provide long-term resupply potential,  and  sufficient mass exists for resupply judging by the far-IR excess reported by \citet{green13c}.  However, the size and mass of such an envelope plus disk is constrained by (unresolved) submillimeter observations to 0.09 M$_{\odot}$ within 650 AU or less (M. Dunham, 2016, priv. comm.).  We therefore also conclude that direct mass transfer between the disks also could not lead to repeated accretion bursts on orbital timescales.

In summary, while the FUor-binary connection remains unclear, the original V1057 Cyg outburst could have been triggered by a binary interaction torquing the 10 AU circumprimary disk.  However, timescale and mass reservoir considerations indicate that FUor outbursts cannot be expected to recur during the next several binary orbits for a system with such a short orbital period.

\acknowledgements

We thank Colette Salyk, Neal Evans, and Yao-Lun Yang for helpful discussions of our results.
This work was supported by a NASA Keck PI Data Award to A. Kraus, administered by the NASA Exoplanet Science Institute. AWM was supported through Hubble Fellowship grant 51364 awarded by the Space Telescope Science Institute, which is operated by the Association of Universities for Research in Astronomy, Inc., for NASA, under contract NAS 5-26555.  Data presented herein were obtained at the W. M. Keck Observatory from telescope time allocated to NASA through the agency's scientific partnership with Caltech and UC. The Observatory was made possible by the generous financial support of the W. M. Keck Foundation. The authors wish to recognize and acknowledge the very significant cultural role and reverence that the summit of Mauna Kea has always had within the indigenous Hawaiian community. We are most fortunate to have the opportunity to conduct observations from this mountain.

\bibliographystyle{../papers/apj.bst}
\bibliography{../papers/krausbib,dissbib.bib}

\end{document}